\documentclass[10pt, a4paper, conference, compsocconf]{IEEEtran}                                                          

\IEEEoverridecommandlockouts                              

\usepackage{graphicx} 
\usepackage{amsmath} 
\usepackage{amssymb}  
\usepackage{color,xcolor,colortbl}
\usepackage{listings}
\usepackage{float}
\usepackage{multirow}

\newcommand{\X}{$\times$}

\definecolor{Gray}{gray}{0.9}

\ifCLASSOPTIONcompsoc
    \usepackage[caption=false, font=normalsize, labelfont=sf, textfont=sf]{subfig}
\else
\usepackage[caption=false, font=footnotesize]{subfig}
\fi

\title{\LARGE \bf
A Longitudinal Analysis of a Social Network of Intellectual History}

\begin{document}

\author{
\IEEEauthorblockN{Cindarella Petz, Raji Ghawi and J\"urgen Pfeffer}
\IEEEauthorblockA{
{Bavarian School of Public Policy}\\
{Technical University of Munich, Munich, Germany} \\
Email: \{cindarella.petz, raji.ghawi, juergen.pfeffer\}@tum.de}
}


\maketitle

\begin{abstract}
The history of intellectuals consists of a complicated web of influences and interconnections of philosophers, scientists, writers, their work, and ideas. How did these influences evolve over time?
Who were the most influential scholars in a period? 
To answer these questions, we mined a network of influence of over 12,500 intellectuals, extracted from the Linked Open Data provider YAGO. We enriched this network with a longitudinal perspective, and analysed time-sliced projections of the complete network differentiating between within-era, inter-era, and accumulated-era networks. We thus identified various patterns of intellectuals and eras, and studied their development in time. 
We show which scholars were most influential in different eras, and who took prominent knowledge broker roles. 
One essential finding is that the highest impact of an era's scholar was on their contemporaries, as well as the inter-era influence of each period was strongest to its consecutive one. Further, we see quantitative evidence that there was no re-discovery of Antiquity during the Renaissance, but a continuous reception since the Middle Ages.

\end{abstract}

\section{INTRODUCTION}
\label{sec:introduction}
``No self is of itself alone", wrote Erwin Schr\"{o}dinger in 1918 \cite{moore:1994} and noted, ``It has a long chain of intellectual ancestors". 
The history of intellectuals is comprised of a myriad of such long chains, embedded in a tapestry of competing influences of ``ageless" ideas, which - in the words of the French scholar Bonaventura D'Argonne in 1699 - ``embrace [...] the whole world" \cite{grafton:2009}. 

To understand the dynamics of influence and spread of ideas through history, the embeddness and interconnections of scholarship should be taken into account.
A network approach offers to identify the most influential scholars via their positions in a network of intellectual influence through the history. 
This allows to study their social relations \cite{Wasserman:1994,Hennig:2012,Otte:2002}, and to provide deep insights into the underlying social structure. 







A recent study by Ghawi et al. \cite{Ghawi:2019c} addressed the analysis of such a social network of intellectual influence incorporating over 12,500 scholars from international origins since the beginning of historiography. 
In this paper, we build upon \cite{Ghawi:2019c}, and extend the analysis of that network by incorporating a temporal dimension. We analyze the network of scholars dependent to their time, adding a longitudinal perspective on how scholars formed networks. 
As such, we opt for an inclusive, global perspective on the history of intellectuals. 
This perspective of a vast longitudinal global network of intellectuals answers to recent discussions on not-global-enough research within intellectual history \cite{haakonssen:2017}. We thus attempt to go beyond the traditional ``master narratives" \cite{ganger:2013} of a Western European centrist view on intellectual history \cite{subrahmanyam:2017}. 
The goal of this paper is not only to understand how the influence relations among scholars evolved over time, but also to get deep insights on their influence on historical periods.
The questions we seek to answer are:
\begin{itemize}
\item How did these influence networks evolve over time?
\item Who were the most influential scholars in a period?  
\item Which patterns of influence did emerge?
\end{itemize}

To answer these questions, we analyze the evolution of influences in time in order to identify periods and scholars, who stand out.


The contributions of this paper are:
\begin{itemize}
\item We incorporate a longitudinal perspective on the social network analysis of intellectuals based on a global periodization of history.
\item We identify patterns of influence, and their distribution in within-, inter-, and accumulated-era influence networks.
\item We identify influence signatures of scholars and eras.
\item We identify scholars with various knowledge broker roles.
\end{itemize}


This paper is organized as follows.
Section \ref{sec:realtedwork} reviews related works. 
In Section \ref{sec:data}, we briefly outline the data set's characteristics and pre-processing.
Section \ref{sec:analysis} presents the network analysis of the entire network, and its time-sliced projections into partial influence networks (within-era, inter-era, and accumulated-era), featuring their basic network metrics, degree distribution and connectivity. 
In section \ref{sec:patterns}, we identify different influence patterns of scholars and eras.
Section \ref{sec:brokerage} is devoted to the longitudinal analysis of brokerage roles in scholars.

\section{RELATED WORK}
\label{sec:realtedwork}
The term of intellectual history combines a plethora of approaches on discourse analysis, evolution of ideas, intellectual genealogies, and the history of books, various scientific disciplines, political thought, and intellectual social context \cite{wickberg:2001, gordon:2013}.
These studies are usually limited to certain regions or time spans as a trade-off for thorough comparative and textual analysis. 
Endeavors to write a ``Global Intellectual History" \cite{moyn:2013} were criticized for focusing on the more well-known intellectual thinkers despite including a transnational comparative perspective \cite{subrahmanyam:2015}. 

Network Methodologies allow to analyze intellectual history and as such the history of intellectuals as big data encompassing time and space with a focus on their inter-connections. 
So far, computational methods were used in the study of communication networks of the \textit{respublica litteraria}, in which various studies modeled the Early Modern scholarly book and letter exchanges as networks. Among the first was ``Mapping the Republic of Letters" at Stanford University in 2008 \cite{edelstein:2017}. More recent studies incorporated a temporal perspective on these epistolary networks \cite{vugt:2017}.

A recent study \cite{Ghawi:2019c} proposed to study the entire history of intellectuals with the means of a network approach. 
This
paper defined the most influential as those with the longest reaching influence (influence cascades), and identified as such Antique and Medieval Islam scholars, and as the one with the most out-going influences, Karl Marx.
In this paper, we extend this analysis by incorporating a temporal dimension, in order to establish a deeper insight on how these influences evolved in time.

Much research has been devoted to the area of longitudinal social networks \cite{Newcomb:1961,Huisman:2003,Snijders:2010,holme:2019}. 
Longitudinal network studies aim at understanding how social structures develop or change over time usually by employing panel data \cite{Hennig:2012}.
Snapshots of the social network at different points in time are analyzed in order to explain the changes in the social structure between two (or more) points in time, in terms of the characteristics of the actors, their positions in the network, or their former interactions. 

In this paper, we do not use the classical notion of network snapshot, which is a static network depicted at a given point in time. Rather, we split the time span (i.e., the history) into consecutive periods (eras), and embed the network nodes (actors) into the eras in which they lived. This way, the micro-level influence among actors can be viewed as a macro-level influence among periods of history. This enables the analysis of the influence network within each era, between different eras, and in an accumulative manner.


In its core, a network of scholarly influence is a citation network, answering to who is influenced by whom \cite{batagelj:2014}.
However, in citation networks the influence is \emph{indirect}, as the relation is originally among documents, from which a social relation among the authors is inferred. In the data set used here, the influence relation is rather \emph{direct} among intellectuals.

\section{DATA}
\label{sec:data}
\subsection{Data Acquisition and Preprocessing}

The source of information used in this paper originated YAGO (Yet Another Great Ontology) \cite{yago:2015}, a pioneering semantic knowledge base that links open data on people, cities, countries, and organizations from Wikipedia, WordNet, and GeoNames.
At YAGO, an influence relation appears in terms of the \verb|influences| predicate that relates a scholar to another when the latter is influenced by the ideas, thoughts, or works of the former. The accuracy of this relation was evaluated by YAGO at 95\%.
We extracted a data set that encompasses all influence relationships available in YAGO, 
using appropriate SPARQL queries that implement social networks mining from LOD techniques \cite{Ghawi:2019}.
The result consisted of 22,818 directed links among 12,705 intellectuals, that made up the nodes and edges of our target social network of influence.
In order to incorporate a time dimension to our analysis, we extracted birth and death dates of each scholar. 
Some scholars had missing birth and/or death dates, which we deduced by subtracting 60 years from the death date, and vice versa, up to the symbolic year of 2020.
When both dates were missing, we manually verified them. During this process, we had to remove some entities, as they did not correspond to intellectuals. Those were either 1) concepts, e.g., `German\_philosophy' and `Megarian\_school', 2) legendary characters, e.g., `Gilgamesh' and `Scheherazade', or 3) bands e.g., `Rancid' and `Tube'. 
To this end, we obtained a new data set of 12,577 actors with complete dates of birth and death.

\begin{figure*}[t!]
\centering
\includegraphics[width=1.0\textwidth]{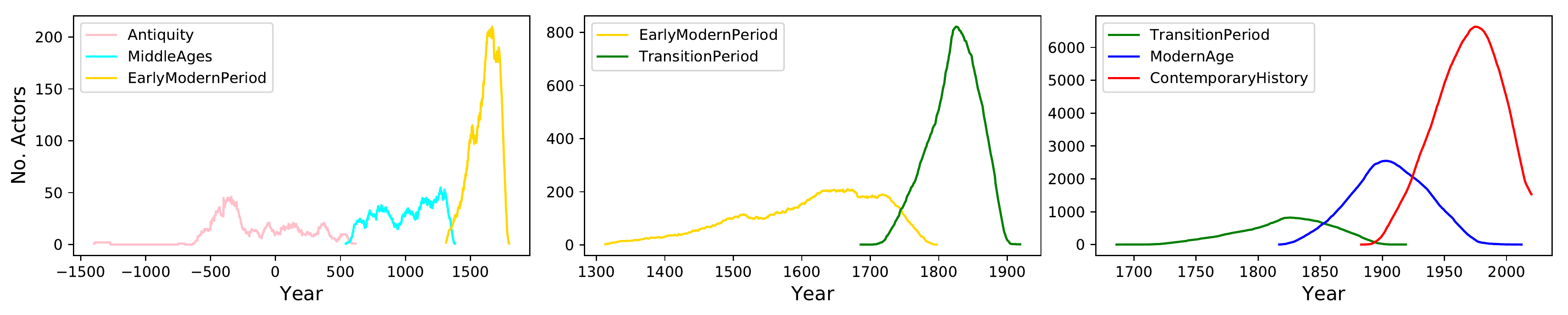}
\caption{Number of actors alive in each year based on their assigned eras.} 
\label{fig:nActors}
\end{figure*}

\subsection{Periodization}

Introducing a longitudinal perspective, we opted for a periodization taking global events into account. Any periodization is a construct of analysis, as each field of research has its own timeline characterizing periods \cite{pot:1999}, which are dependent on different caesura for the respective object of research \cite{osterhammel:2006}. This complicates an overarching longitudinal perspective on a global scale. We used Osterhammel's global periodization \cite{osterhammel:2006} to match the internationality of the scholars, and worked with six consecutive periods (eras): Antiquity (up to 600 AD), Middle Ages (600, 1350), Early Modern Period (1350, 1760), Transitioning Period (1760, 1870), Modern Age (1870, 1945), and Contemporary Period (1945, 2020).

One conceptual challenge was to map actors into eras.
Many actors fit to more than one period's timeline. We opted for a single era membership approach since it is more intuitive and easier to conceive, and reduces the complexity of analysis and computations, while grasping the essential membership to an era of each scholar.
It also offers adequate results when we compare eras, since it avoids redundancy.

In order to assign a single era to an actor, we used the following method: We assumed that scholars would not be active in the first 20 years of their lives. Therefore, we calculated the mid point of the scholar's lifespan ignoring the first 20 years of their age, then we assigned the era in which this mid point occurs. 

After this initial assignment process, we verified the global validity of assignments by counting the number of influence links from one era to another. We observed that there were some reverse links of eras, i.e., an influence relation from an actor in a recent era towards an actor assigned to an older era. Those anomaly cases (about 200) were basically due to:
\begin{itemize}
\item Errors in dates: 
\begin{itemize}
\item some dates were stated in the Hijri calendar, instead of the Gregorian calendar, and
\item some dates were BC and missing the negative sign.
\end{itemize}

\item Errors in direction of the relationship: 
source and target actors were wrongly switched.
\item Inappropriate era-actor assignments.
\end{itemize}

The anomalies due to errors have been manually corrected.
The cases of inappropriate assignment were technically not erroneous. This usually happened when the influencer lived much longer than the influenced, elevating the influencer's period into a more recent one. 
We solved this by iteratively re-assigning either the influencer backward to the era of the influenced, or the influenced forward to the era of the influencer. As a result, each actor is assigned to exactly one era, such that no reverse links of eras exist. The final cleaned dataset consists of 22,485 influence links among 12,506 actors.

\section{ANALYSIS}
\label{sec:analysis}

Figure \ref{fig:nActors} shows each era's continuous density of scholars based on their lifespan.

With scholars embedded in their respective eras, the entire influence network can be time-sliced: we projected it into several partial networks based on the source era (of the influencer) and target era (of the influenced scholar). When the source and target eras are the same, we called the partial network a \emph{within-era} influence network. When the source and target eras are different, we called the partial network an \emph{inter-era} influence network. There are no reverse links from a later era to a previous one due to pre-processing.

After time-slicing the whole network, we received
six within-era networks corresponding to all the six eras, and 15 inter-era networks, corresponding to all chronologically ordered but not necessarily consecutive pairs of different eras. Additionally, we constructed six \emph{accumulated-era} influence networks of all scholars living up to and including the target era.


\begin{figure}[t]
\centering
\includegraphics[width=\linewidth]{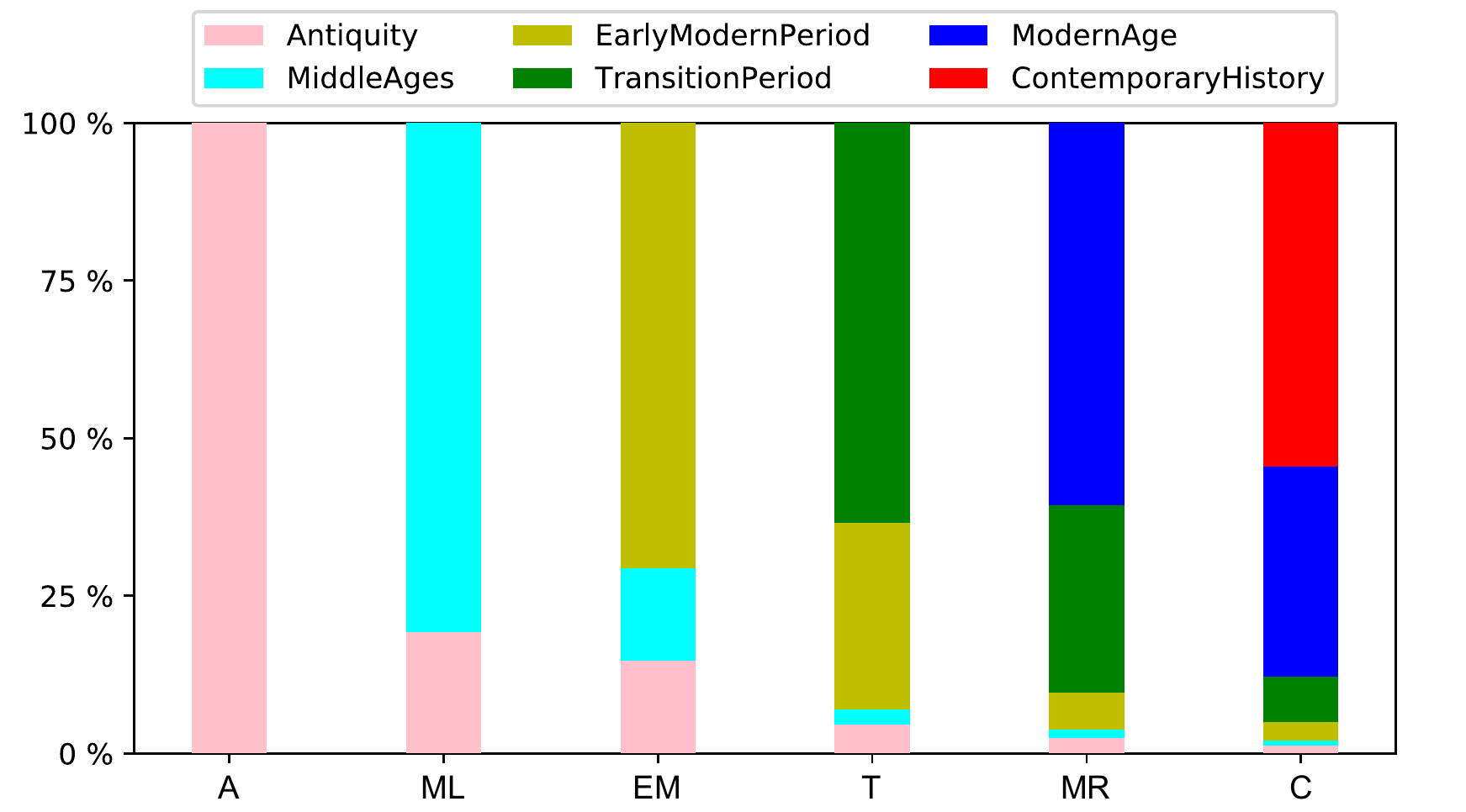}
\caption{Percentage of received influences in each era.}
\label{fig:inflc_pc}
\end{figure}

Figure \ref{fig:inflc_pc} shows the proportion of influence links among all pairs of eras. 
There, we can make already two major observations for inter- and within-era influence relations:
For one, the highest fraction of influence received by scholars of each era come from its own era. This means that the internal impact of any era is in general higher than its external impact. In absolute numbers, the vast majority of links occur within the Contemporary era, followed by links from Modern Age to Contemporary, and within Modern Age, which is clearly owed to the increased amount of scholars in these periods.\\
The inter-era influences of each period is strongest to its consecutive period. As our earliest period, Antiquity receives only influence links from itself, whereas the influence received in the Middle Ages are 82\% internal, and 18\% from Antiquity. Subsequently, the amount of the within-era influence shrinks throughout the consecutive periods, but still remains the biggest influence. 
Noteworthy here is the high proportion 
of influences of Antiquity on the Early Modern Period, which represents their increased reception during the Renaissance. 
However, the proportionately many links of Antiquity to the Middle Ages 
reassert the shift in historical research that the Renaissance did not ``re-discover" Antiquity, but was received before in the Middle Ages as well \cite[p. 3-4]{fejfer:2003}.

\subsection{Within-Eras Influence Networks}
\label{sec:within-eras}


In the following, we analysed the six \emph{within-era} influence networks, which represent the internal impact of an era. We 
extracted the following metrics, as shown in Table \ref{tab:within-eras}:
\begin{itemize}
\item Number of nodes $N$, and edges $E$, and density $D$.
\item Average out-degree (= avg. in-degree due to the properties of a directed graph)
\item Max. in-degree, max. out-degree, and max. degree.
\item WCC: number of weakly connected components.
\item LWCC: size of the largest weakly connected component.
\item SCC: number of strongly connected components, when the number of nodes is $> 1$).
\item Reciprocity and transitivity.
\end{itemize}

\begin{table}[ht]
\scriptsize
\centering
\caption{Metrics of Within-Era Networks}
\label{tab:within-eras}    
\begin{tabular}{l | r r r r r r}
Era &  A & ML & EM & T & MR & C \\
\hline
$N$         & 219 & 303 & 610 & 761 & 2102 & 6081 \\
$N/A$     & 82\% & 86\% & 81\% & 70\% & 73\% & 85\% \\
$E$         & 327 & 387 & 694 & 927 & 2806 & 7960 \\
\hline
Density & .0068 & .0042 & .0019 & .0016 & .0006 & .0002 \\
avg. out-degree	& 1.49 & 1.28 & 1.14 & 1.22 & 1.33 & 1.31 \\
max in-degree   & 12 & 9  & 17 & 27 & 21 & 26 \\
max out-degree	& 20 & 16 & 23 & 32 & 68 & 58 \\
max degree	    & 32 & 20 & 32 & 41 & 73 & 58 \\
\hline
WCC & 11 & 21 & 94 & 108 & 208 & 582 \\
Largest WCC & 179 &	233 & 245 & 436 & 1495 & 4379 \\
 & 82\% & 77\% & 40\% & 57\% & 71\% & 72\% \\
SCC & 	0 & 2 & 6 & 8 & 31 & 38 \\
\hline
Reciprocity & 0 & 0.005 & 0.023 & 0.028 & 0.036 & 	0.014 \\
Transitivity & 0.064 & 0.066 & 0.071 & 0.042 & 	0.029 &	0.017 \\
\end{tabular}
\end{table}

We included $\frac{N}{A}$ in Table \ref{tab:within-eras} in order to contain that the number of nodes $N$ in a within-era network could be less than the number of actors of that era $A$. This is owed to the fact that not all scholars of an era necessarily participated in its within-era influence network. Some scholars influenced or were influenced by actors of different eras only. However, around 80\% of scholars in each era were active in these within-era networks. The highest value of 86\% of the Middle Ages refer to their relative self-containment as an era, as well as the lowest value in the Transitioning period of 70\% to its high out-going influences.



Over all eras, the amount of nodes and edges steadily increased, while the density of networks decreased.
On average, the out-degree is relatively stable around 1.25, where the highest value of 1.5 occurs in Antiquity, and the lowest of 1.14 in the Early Modern period. When we compare the evolution of the max. out-degree in time, we find that the expected continuous increase did not always hold due to two exceptionally high observations at Antiquity and Modern Age.
Mutual ties among contemporaries were in general very low. We can report none in Antiquity, and only one in the Middle Ages between Avicenna and Al-Bīrūnī. In the Early Modern period, eight mutual relations were observed, including e.g. Gottfried Leibniz (1646-1716) and David Bernoulli (1700-1782), whereas 13 mutual relations in the Transitioning period, such as Friedrich Engels (1820-1895) and Karl Marx (1818-1883), or Johann Goethe (1749-1832) and Friedrich Schelling (1775-1854). In the Modern Age, the number of mutual ties increased to 51 (e.g. Jean-Paul Sartre (1905-1980) and Simone de Beauvoir (1908-1986)); and to 54 in the Contemporary period.

\begin{figure}[H]
\centering
\includegraphics[width=\linewidth]{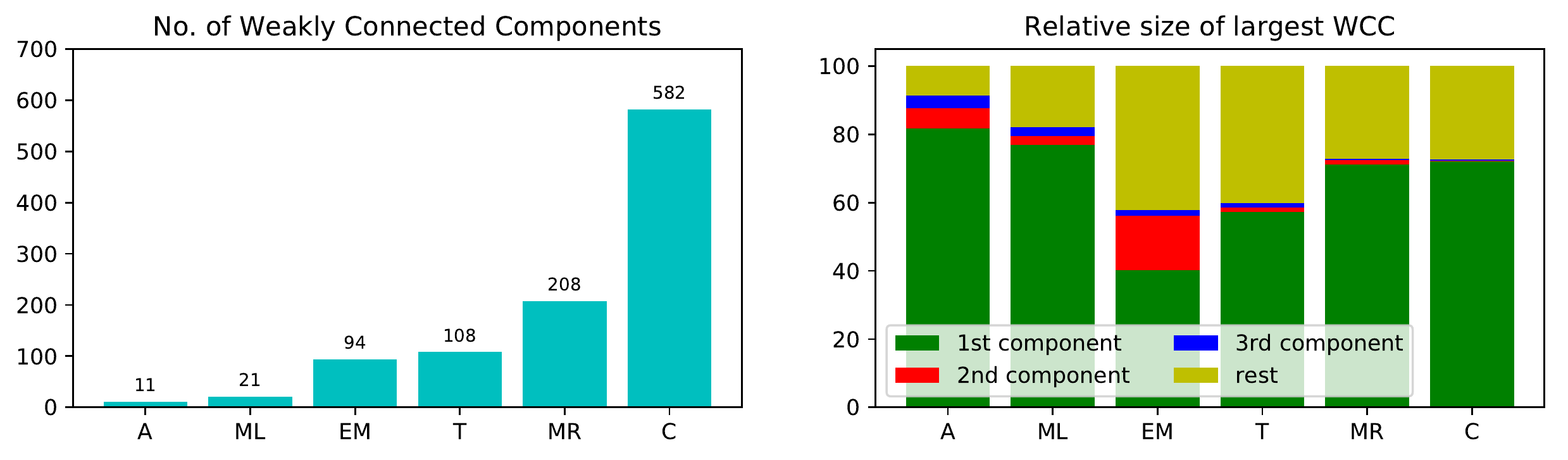}
\caption{Weakly connected components in within-era influence networks}
\label{fig:within-WCC}
\end{figure}

Figure \ref{fig:within-WCC} shows the number of weakly connected components (WCCs) in the within-era networks of each era, and the relative size of the largest ones w.r.t the whole corresponding network. 
The number of WCCs increased gradually over the consecutive eras. In general, the networks consisted of one giant component, which encompassed the majority of nodes, while the rest of components were relatively smaller. 
This was particularly developed in Antiquity and Middle Ages, where the giant component constitute of 82\% and 77\% of the nodes, while the second largest were at 6\% and 3\%, respectively. 
The Early Modern period constitutes an exception to this giant component rule: the largest one was at 40\% only, and the second largest at 16\%. Looking at their composition, the first consisted of natural scientists, mathematicians, and philosophers, such as Descartes, Newton, and Leibniz, while the smaller one compromised of artists and painters, such as Rembrandt and Raphael.
The single giant component phenomenon appeared again in subsequent eras. For instance, in the Transitioning period, there were 108 WCCs, where the largest two incorporated 57\% and 1.3\% of the nodes.
In Modern and Contemporary age, the largest components comprised about 70\% of nodes.

\begin{table}[t]
\scriptsize  
\centering
\caption{Top 5 actors, per era, based on out-degree in within-era influence networks.}
\label{tab:within-top5outdeg}
\begin{tabular}{l r | l r | l r}
\hline
\rowcolor{Gray}
Antiquity &     & MiddleAges & & EarlyModern & \\
\hline
Plato      & 20 & Avicenna   & 16 & John Locke     & 23 \\ 
Aesop      & 13 & Muhammad   & 11 & René Descartes & 22\\ 
Pythagoras & 10 & Al-Ghazali & 11 & Isaac Newton   & 15\\
Plotinus   & 10 & Banū Mūsā  &  8 & Hugo Grotius   & 13\\
Euhemerus  & 10 & J. S. Eriugena & 8 & Leibniz & 11\\
\hline \hline 
\rowcolor{Gray} Transition & & Modern & & Contemporary &  \\
\hline 
Goethe        & 32 & Nietzsche      & 68 & Vladimir Nabokov & 58 \\
Hegel         & 29 & Jules Verne    & 35 & Friedrich Hayek  & 50 \\
Lord Byron    & 24 & Henri Bergson  & 35 & Richard Pryor    & 50 \\
Immanuel Kant & 22 & Leo Tolstoy    & 24 & Jacques Derrida  & 48 \\
von Schelling & 17 & Edmund Husserl & 22 & Michel Foucault  & 47 \\
\hline
\end{tabular}
\end{table}

\emph{Who was most influential on their contemporaries?}
Table \ref{tab:within-top5outdeg} lists the top five scholars per era based on their out-degree in the within-era influence networks. The highest within-era out-degree over all times was achieved by Friedrich Nietzsche (1844-1900) of the Modern Age with 68 outgoing influence links to other scholars of his era.

\subsection{Inter-Era Influence Networks}
\label{sec:inter-era}

\emph{Inter-era influence} networks are partial networks where the source era is preceding the target era. We interpreted these networks as bipartite, as the actors belong to different groups, the source era and the target era. Therefore, only edges between nodes sets are possible. 

\begin{table}[H]
\centering
\caption{Metrics of inter-eras influence networks}
\label{tab:inter-eras}
\scriptsize 
\setlength\tabcolsep{3pt}
\begin{tabular}{l | r r | r r | c | c c | c c |}
source $\rightarrow$ & N & E & $N_s$ & $N_t$ & D & \multicolumn{2}{c|}{in-degree} &  \multicolumn{2}{c|}{out degree} \\
\cline{7-10}
 target  &   &   &       &       &         & avg & max & avg  & max \\
\hline 

A $\rightarrow$	MA	&  82 &  87 & 38 &  44 & .052 & 1.98 & 7  & 2.29 & 12	\\
A $\rightarrow$	EM	& 117 & 145 & 46 &  71 & .044 & 2.04 & 7  & 3.15 & 19	\\
A $\rightarrow$	T	&  66 &  66 & 29 &  37 & .062 & 1.78 & 5  & 2.28 & 11	\\
A $\rightarrow$	MA	& 101 & 114 & 42 &  59 & .046 & 1.93 & 11 & 2.71 & 23	\\
A $\rightarrow$	C	& 169 & 177 & 49 & 120 & .030 & 1.47 & 6  & 3.61 & 46	\\
\hline

ML $\rightarrow$ EM & 149 & 144 & 66 & 83 & .026 & 1.73 & 9 & 2.18 & 21	\\
ML $\rightarrow$ T  &  52 &  36 & 22 & 30 & .055 & 1.20 & 5 & 1.64 &  6	\\
ML $\rightarrow$ MR &  77 &  62 & 27 & 50 & .046 & 1.24 & 4 & 2.30 & 12	\\
ML $\rightarrow$ C  & 146 & 121 & 50 & 96 & .025 & 1.26 & 6 & 2.42 & 34  \\
\hline
EM $\rightarrow$ T	&   392 &   432 &   159 &   233 & .012 & 1.85 & 16 & 2.72 & 24	\\
EM $\rightarrow$ MR	&   262 &   269 &   101 &   161 & .016 & 1.67 & 13 & 2.66 & 15	\\
EM $\rightarrow$ C	&   437 &   432 &   125 &   312 & .011 & 1.38 &  7 & 3.46 & 35	\\
\hline
T $\rightarrow$ MR	& 1,111 & 1,373 &   436 &   675 & .005 & 2.03 & 19 & 3.15 &  53 \\
T $\rightarrow$ C	&   888 & 1,041 &   212 &   676 & .007 & 1.54 &  9 & 4.91 & 112 \\
\hline
MR $\rightarrow$ C	& 3,817 & 4,885 & 1,271 & 2,546 & .002 & 1.92 & 17 & 3.84 &  78 \\	
\hline
\end{tabular}

\end{table}

Table \ref{tab:inter-eras} shows the metrics for those inter-era influence networks. 
In general, each era had the most links with its consecutive era, and additionally with the Contemporary period's scholars. Exception to this rule was Antiquity, which saw its first peak with the Early Modern period relating to Renaissance interests. 
Their densities were again decreasing through the combinations, except for those periods that had less links to other periods, such as the Middle Ages to the Transitioning period.\\

\begin{table}[b]
\scriptsize
\centering
\caption{Top scholars with highest out-degree in the inter-era networks}
\label{tab:inter-top}
\begin{tabular}{l | lr | lr  }
s $\rightarrow$ t  & \multicolumn{2}{c|}{First Rank}
& \multicolumn{2}{c}{Second Rank}
\\
\hline
A $\rightarrow$ ML & Aristotle & 12 & Augustine of Hippo & 6 \\
A $\rightarrow$ EM & Aristotle & 19 & Plato     & 14 \\
A $\rightarrow$ T  & Aristotle & 11 & Plato     & 9  \\
A $\rightarrow$ MR & Plato     & 23 & Aristotle & 16 \\
A $\rightarrow$ C  & Aristotle & 46 & Plato     & 32 \\
\hline
ML $\rightarrow$ EM & Ibn Tufail & 21 & Thomas Aquinas & 9 \\
ML $\rightarrow$ T & Petrarch & 6 & Dante Alighieri & 5 \\
ML $\rightarrow$ MR & Dante Alighieri & 12 & Thomas Aquinas & 11 \\
ML $\rightarrow$ C & Thomas Aquinas & 34 & Dante Alighieri & 10 \\
\hline
EM $\rightarrow$ T  & J. J. Rousseau & 24 & Shakespeare & 21 \\
EM $\rightarrow$ MR & Baruch Spinoza & 15 & Shakespeare & 15 \\
EM $\rightarrow$ C  & Shakespeare    & 35 & David Hume  & 25 \\
\hline
T $\rightarrow$ MR & Immanuel Kant & 53 & Karl Marx & 43 \\
T $\rightarrow$ C  & Karl Marx    & 112 & Hegel     & 67 \\
\hline
MR $\rightarrow$ C & Nietzsche & 78 & Martin Heidegger & 73 \\
\end{tabular}
\end{table}

\emph{Which scholar influenced a successive era the most?}
Table \ref{tab:inter-top} shows the scholars with the highest degrees in the inter-era networks. Noteworthy here is Karl Marx, who had the highest out-degree over all times from the Transitioning period to the Contemporary age, followed by modern philosopher Friedrich Nietzsche and Martin Heidegger on Contemporary scholars.

\subsection{Accumulative Influence Networks}
\label{sec:accumulative}
For each era, we constructed an accumulative influence network of all influence links among scholars who lived up to and including that era. We performed essential social network analysis on these six \emph{accumulated-eras} networks, which combine the internal and external impact of eras. The final network of Contemporary Age is the same as the complete network over all periods \cite{Ghawi:2019c}.

\begin{figure}
\centering
\includegraphics[width=\linewidth]{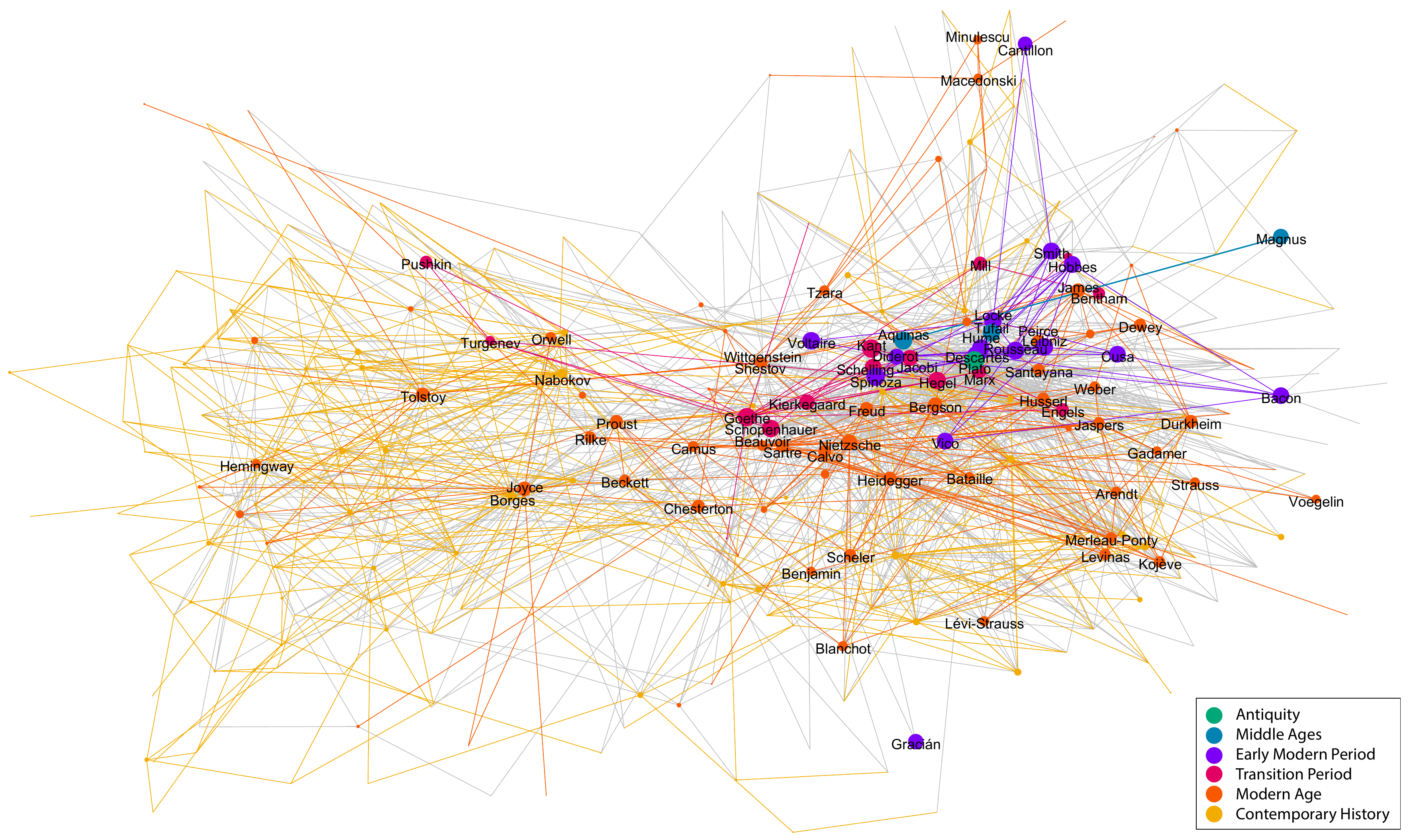}
\caption{Network of the most influential actors with at least 10 out-going influences. Node size = proximity prestige, node color = era, links within an era are colored with the color of the era, the other links are gray.}
\label{fig:network}
\end{figure}

Fig. \ref{fig:network} shows the best connected scholars for each era, that influence at least 10 others, in the final accumulated network. We clearly see two joined networks of hubs. The right part is very diverse in terms of including different eras and different fields such as philosophy, theology and science scholars. The left part consists mainly of writers since the long 19. Century (1789-1914); Alexander Pushkin (1799-1837) is one of the eldest nodes there. This writers' network shows little diversity to other historical periods and consists mostly of Modern and Contemporary age writers. 
That writers are less connected to the philosophy, theology, and science scholars shows that these groups referenced themselves more consistently.

\begin{table}[b]
\scriptsize
\centering
\caption{Metrics of Accumulative-Era Networks}
\label{tab:accumulative-eras}    
\begin{tabular}{l | r r r r r r}
 
Era &  
 A & ML & EM & T & MR & C \\
\hline
$N$         & 219 & 552 & 1,227 & 2,141 & 4,697 & 12,506 \\
$E$         & 327 & 801 & 1,784 & 3,245 & 7,869 & 22,485 \\
$N_{src}$   &  54 & 155 &   388 &   677 & 1,501 &  3,890 \\
$N_{inner}$ &  71 & 178 &   353 &   597 & 1,331 &  3,080 \\
$N_{sink}$  &  94 & 219 &   486 &   867 & 1,865 &  5,536 \\
\hline
Density        & .0068 & .0026 & .0012 & .0007 & .0004 & .0001  \\
avg. out-degree	& 1.49 & 1.45 & 1.45 & 1.5 & 1.68 & 1.80 \\
max in-degree   &  12 & 16 & 26 & 38 &  48 & 48 \\
max out-degree	&  20 & 24 & 41 & 52 &  75 & 158 \\
max degree	    &  32 & 36 & 50 & 60 & 116 & 196 \\
\hline
WCC & 11 & 30 & 110 & 211 & 390 & 817 \\
Largest WCC & 179 &	441 & 797 & 1513 & 3550 & 10192 \\
 & 82\% & 80\% & 65\% & 71\% & 76\% & 81\% \\
SCC & 	0 & 2 & 8 & 16 & 47 & 85 \\
\hline
Reciprocity & 0 & 0.002 & 0.010 & 0.014 & 0.019 & 0.011  \\
Transitivity & 0.064 & 0.067 & 0.064 & 0.056 & 	0.039 &	0.021 \\
\end{tabular}
\end{table}

Table \ref{tab:accumulative-eras} shows the metrics of accumulated-era networks. 
Regarding node degrees change over consecutively accumulated eras, we observe that at all eras the maximum out-degree is greater than the maximum in-degree. 
Moreover, those maximum degrees continuously increase over eras, in contrast to within-era networks.
The average out degree changes slightly over time, taking its lowest value of 1.45 at Middle Ages, and highest value of 1.8 at Contemporary age. Noteworthy is the drastic collapse of the largest Weak Component in the Early Modern period, which was steadily rising since.

\begin{figure}[t]
\centering
\includegraphics[width=\linewidth]{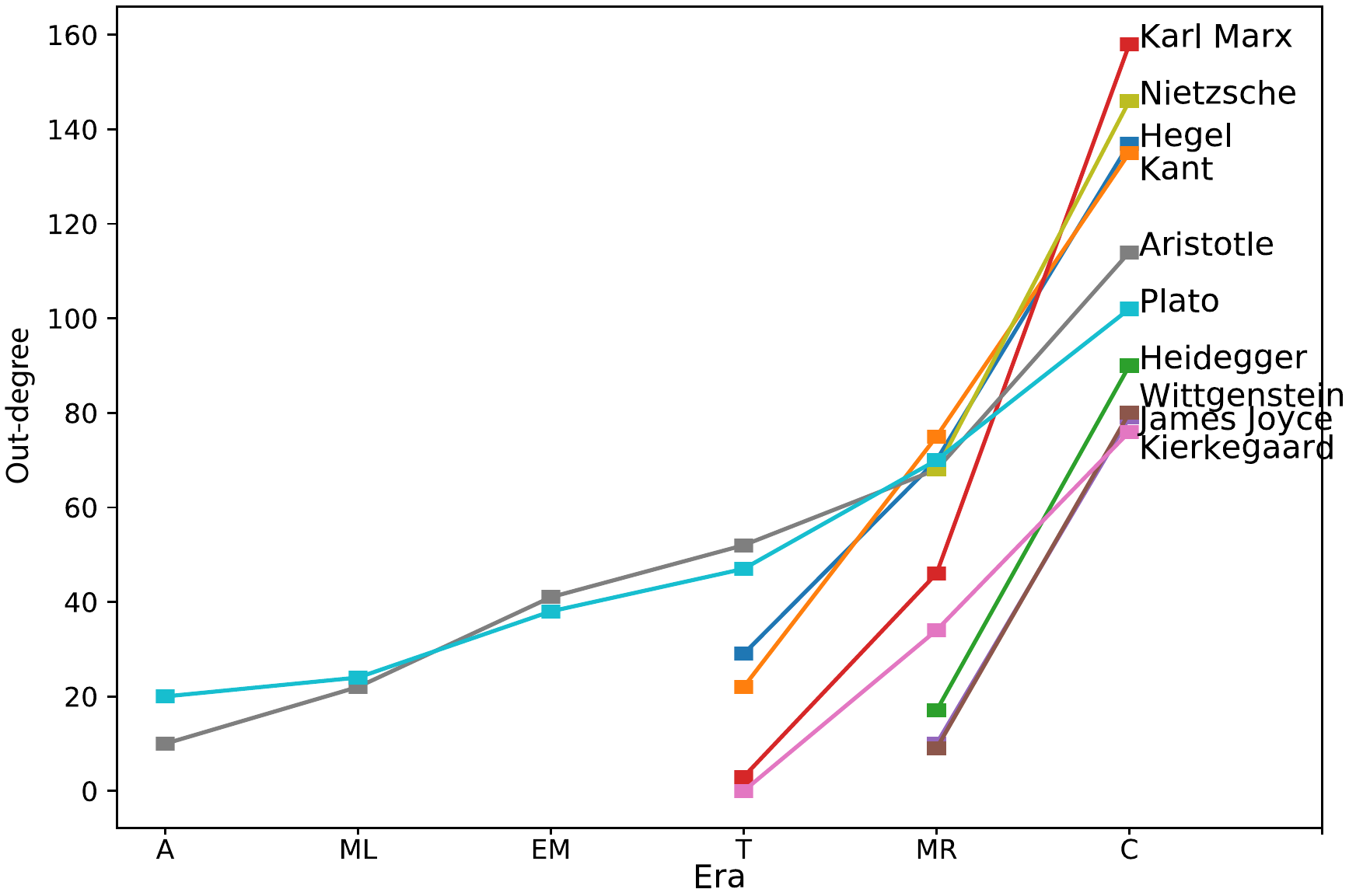}
\caption{Top 10 of the most influential intellectuals of the complete network based on their out-degree, and their progression in the accumulated-era networks.}
\label{fig:top10}
\end{figure}

\emph{Who was the most influential intellectual in an era?} 
Figure \ref{fig:top10} shows the evolution of the ten most influential scholars in the complete network based on their out-degree progressing in the accumulative networks.\\ 
The top two ranks of the most prolific scholars were consistently taken over by Antique philosophers Plato, and Aristotle (who among contemporaries was only in rank 6) 
Contemporary scholars came on third rank in the Middle Ages (Avicenna), in the Early Modern period (Ibn Tufail, John Locke, René Descartes), and in the Transitioning period (John Locke, Johann Goethe). This changed in the Modern Age, when Transitioning period scholars Immanuel Kant and Hegel took the first ranks. Aristotle still remained in the top five. The highest out-degree over all times is observed at Contemporary age, where Karl Marx had 158 outgoing influence links to other scholars of all eras, followed by Nietzsche, Hegel and Kant.



\section{Patterns of Influence over Eras}
\label{sec:patterns}

In this section, we study the influence patterns of scholars over eras. We construct influence signatures based on how much on average a scholar influenced an era, and which patterns of directed influences characterize an era.

\subsubsection{Influence Power of Scholars}
For each scholar, we construct their influence signature as a sequence of their influence links towards each era starting from their own.
For example, the influence signature of Aristotle was $[10, 12, 19, 11, 16, 46]$, which meant, he had 10 influence links within Antiquity, 12 links towards the Middle Ages, etc. 
Using those signatures, we define the \emph{longitudinal influence power} of a scholar as the average of their influence signature.
A scholar would have a high influence power when he has (1) a high number of influence links (2) over all or many eras.
In contrast, having few influence links over several eras, or many links over few eras would give low value of this influence power measure. For example, with an average around 19 both Aristotle and Shakespeare had similar influence powers. In absolute numbers, Aristotle had almost twice the number of Shakespeare's influence links (114 to 73, respectively). While Aristotle influenced all 6 eras, and Shakespeare only 4, the ratio of the links per era decreased for Aristotle, resulting in their similar influence powers.
This measure provides an indicator of the influence power of an intellectual throughout history, and combines both the intensity and the diversity of influence.

It also allows us to compare scholars from different eras. Table \ref{tab:power} shows the top 5 scholars based on the longitudinal influence power.
Here, Aristotle, Thomas Aquinas, William Shakespeare, Karl Marx, Friedrich Nietzsche and the writer Vladimir Nabokov (1899-1977) are identified by their influence power as the most influential intellectuals of their respective periods. The highest longitudinal influence powers over all times had Nietzsche (73), followed by Nabokov (58) and Marx (52).

\begin{table}[ht]
\scriptsize 
\centering
\setlength\tabcolsep{3pt}
\caption{Top 5 actors based on the longitudinal influence power.}
\label{tab:power}
\begin{tabular}{l r || l r || l r}
\hline
\rowcolor{Gray}
Antiquity &     & MiddleAges & & EarlyModern & \\
\hline
Aristotle          & 19.0 & Thomas Aquinas  & 12.6 & William Shakespeare & 18.2 \\
Plato              & 17.0 & Dante Alighieri & 6.0  & Baruch Spinoza      & 14.8 \\
Augustine of Hippo &  6.0 & Ibn Tufail      & 5.8  & René Descartes      & 14.0 \\
Plotinus           &  4.7 & Avicenna        & 4.6  & John Locke          & 13.0 \\
Heraclitus         &  4.2 & Al-Ghazali      & 3.6  & David Hume          & 12.5 \\
\hline \hline 
\rowcolor{Gray} Transition &  & ModernAge & & Contemporary & \\
\hline 
Karl Marx & 52.6   & Friedrich Nietzsche & 73.0  & Vladimir Nabokov & 58.0 \\
Hegel     & 45.7   & Martin Heidegger    & 45.0  & Friedrich Hayek  & 50.0 \\
Immanuel Kant & 45.0 & Ludwig Wittgenstein & 40.0  & Richard Pryor    & 50.0 \\
Søren Kierkegaard & 25.3 & James Joyce         & 39.5  & Jacques Derrida  & 48.0 \\
Fyodor Dostoyevsky & 23.0 & Sigmund Freud       & 32.0  & Michel Foucault  & 47.0 \\
\hline 
\end{tabular}
\end{table}


\subsubsection{Influence Patterns}
Which directed influences were most common in an era? We derive to these influence patterns of eras by replacing any non-zero entries by X of the scholar's influence signatures, and aggregate all occurrences of each pattern for each era. 
We thus ignore the actual values of influence (intensity), but keep the temporal effect (diversity).
For example, the influence pattern $[ X, 0, \cdots, 0]$ means that the scholarly influences goes to the first (own) era only, with no influence on other eras. 
The pattern $[ X, X, \cdots, X]$ signifies that the influence is distributed over all applicable eras, regardless of the actual values.
Table \ref{tab:infl_patterns} gives the top patterns of each era with the pattern's frequency of occurrence with regard to the respective era.

\begin{table}[ht]
\scriptsize 
\centering
\caption{Top frequent influence patterns of eras (from left to right)}
\label{tab:infl_patterns}    
\begin{tabular}{l| c c c c c c | r }
& A & ML & EM & T & MR & C & \\
\hline
\multirow{4}{*}{Antiquity}
  & \X & 0  & 0  & 0 & 0 & 0  & 43\% \\
  & 0  & 0  & 0  & 0 & 0 & \X & 8\% \\
  & 0  & \X & 0  & 0 & 0 & 0  & 7\% \\
  & 0  & 0  & \X & 0 & 0 & 0  & 7\% \\
\hline
\multirow{4}{*}{MiddleAges}
  & & \X & 0  & 0 & 0 & 0  & 56\% \\
  & & 0  & \X & 0 & 0 & 0  & 9\% \\
  & & \X & \X & 0 & 0 & 0  & 7\% \\
  & & 0  & 0  & 0 & 0 & \X & 6\% \\
\hline           
\multirow{4}{*}{EarlyModern}
  & & & \X & 0  & 0  & 0  & 51\% \\ 
  & & & 0  & \X & 0  & 0  & 13\% \\
  & & & 0  & 0  & 0  & \X & 7\% \\
  & & & \X & \X & \X & \X & 7\% \\
\hline                  
\multirow{6}{*}{Transition}
  & & & & \X & 0  & 0  & 35\% \\
  & & & & 0  & \X & 0  & 29\% \\
  & & & & \X & \X & \X & 11\% \\
  & & & & \X & \X &  0 & 9\% \\
  & & & & 0  & 0  & \X & 8\% \\
  & & & & 0  & \X & \X & 7\% \\
\hline           
\multirow{3}{*}{ModernAge}
  & & & & & 0  & \X & 38.8\% \\
  & & & & & \X & 0  & 36.7\% \\
  & & & & & \X & \X & 24.5\% \\
\hline           
Contemporary 
  & & & & & & \X & 100\% \\               
\end{tabular}

\end{table}

For example, for the Middle Ages the most frequent pattern was
$[-, X, 0, 0, 0, 0]$, which represents that 56\% of scholars only influenced contemporaries with no influences on other eras. Over all eras, the most common pattern was of within-era influence, followed by the influence on the consecutive period. Exception to this rule was the Modern period, which had this rule reversed, and an higher influence on the Contemporary period than on its own. 
Since the Early Modern period, the pattern of influencing all successive eras including its own becomes more frequent (from 7\% on), and rising in each successive period. 

\section{Brokerage Role}
\label{sec:brokerage}
\emph{Which roles had scholars in regard to their influence on others?} We look at roles by following the brokerage approach by Gould and Fernandez \cite{gould:1989} by analyzing non-transitive triads, in which a node A has a tie to node B, and B has a tie to node C, but there is no tie between A and C. 
In these triads, B is thought to play a structural role called a \emph{broker}. 

\begin{figure}[H]
\centering
\includegraphics[width=1.0\linewidth]{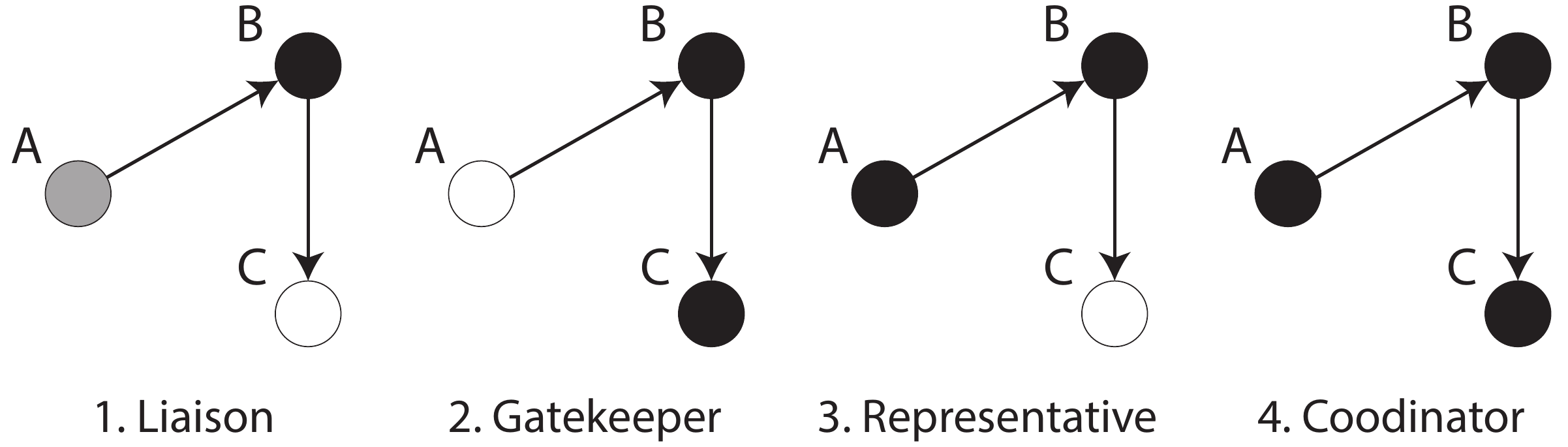}
\caption{Brokerage Roles of the top right node of each triad, adapted from Gould and Fernandez (1989) \cite{gould:1989}.}
\label{fig:brok_roles}
\end{figure}

The possible roles are shown in Figure \ref{fig:brok_roles}, which are adapted from the work of Gould and Fernandez in \cite{gould:1989}, and Everett and Borgatti \cite{everett:2012}.\footnote{The fifth brokerage role, the \emph{Consultant}, where A and C belong to one period, and B belongs to another, is not possible in our network, as we didn't allow reverse influences of a more recent period onto a previous one by pre-processing.}
This allows us to consider to what extent a node's importance is based on joining two nodes that are members of the node's own era, or on joining others outside their group. We interpret nodal membership in groups as eras.

In Table \ref{tab:actor_roles}, we analyse the above described brokerage roles for each period. Over all eras, 23\% of all scholars on average had at least one of the above described brokerage roles. 
Since the Early Modern period, the amount of scholars with exactly one brokerage role stays very stable on about 12-13\%, slightly higher in the Antiquity and Middle Ages.
Both the first and the last of the periods could have a maximum of 2 different brokerage roles, because pre-processing didn't allow reverse links. Therefore, Representative and Liaison brokerage was impossible for Contemporary, as well as Liaison and Gatekeeper brokerage for Antiquity.
Coordinator and Gatekeeper roles represent the scholars importance within their own period. Gatekeeper had inter-period influences, and in turn influenced their contemporaries. The scholars with the highest scores for Gatekeeper in their respective periods were medieval polymath Avicenna (980-1037), Early Modern philosopher René Descartes (1596-1650), and Immanuel Kant (1724-1804), Friedrich Nietzsche (1844-1900), and Michel Foucault (1926-1984). 
The highest scores of Coordinators had Plato, again Avicenna, John Locke (1632-1704), Johann Goethe (1749-1832), again Friedrich Nietzsche, and contemporary horror writer Stephen King (born 1947). As Coordinators, these scholars represented an within-period influence. 
Liaison brokers would have the longest time frame of influence, which includes three successive periods. Highest scores had Dominican friar Thomas Aquinas (1225-1274), Early Modern philosopher Baruch Spinoza (1632-1677), and again Immanuel Kant and Friedrich Nietzsche as Liaisons.
Representatives took the reverse role of an Gatekeeper: They had an within-era influence, that spread to a successive era. Here, Plato, Thomas Aquinas, David Hume (1711-1776), Karl Marx (1818-1883) and Martin Heidegger (1889-1976) stood out.

From Middle to Modern Age, the amount of scholars with all four brokerage roles steadily increased. Noteworthy here were Thomas Aquinas (Middle Ages), Gottfried Leibniz (Early Modern Period), Georg Hegel (Transitioning Period), and Martin Heidegger (Modern Age), who appeared most often in super brokerage roles: They combined Liaison, Gatekeeper, Representative, and Coordinator roles alike in their respective periods. 
Surprisingly though, scholars with 3 brokerage roles were roughly ten times less common than those with all brokerages (compare Table \ref{tab:actor_roles}). 


\begin{table}[H]
\scriptsize 
    \centering
    \caption{Number and fraction of actors taking 1, 2, 3 or 4 roles}
    \label{tab:actor_roles}    
    \begin{tabular}{l| r | r | r | r}
No. of Roles  &   1        &   2        &    3       &       4    \\
\hline                    
Antiquity     &  55 (21\%) &  30 (11\%) &            &            \\
MiddleAges    &  62 (18\%) &  32 (9\%)  &            & 12 (3\%)  \\
EarlyModern   & 101 (13\%) &  51 (7\%)  & 2 (0.3\%) &  38 (5\%)  \\
Transition    & 136 (12\%) &  87 (8\%)  & 6 (0.8\%) &  70 (6\%)  \\
ModernAge     & 363 (13\%) & 269 (9\%)  & 5 (0.7\%) & 200 (7\%)  \\
Contemporary  & 879 (12\%) & 536 (7\%)  &            &            \\
\hline
 overall            & 1,596  & 1,005  & 13    & 320   \\
                    & 12.8\% &  8.0\% & 0.1\% & 2.6\% \\                    
\hline
    \end{tabular}
\end{table}







\section{CONCLUSIONS}
\label{sec:conclusion}

In this paper, we incorporated a longitudinal aspect in the study of the influence networks of scholars. 
First, we extracted their social network of influence from YAGO, a pioneering data source of Linked Open Data. Rigorous pre-processing resulted in a network of 12,705 intellectuals with 22,818 edges, including information on each scholar's era.
We opted for a global approach to the periodization of history, resulting in six consecutive eras to study. 

Our main question was whether we could identify patterns of influence, and their change over time.
Therefore we performed essential network analysis on every time-sliced projection of the entire network in within-era, inter-era, and accumulated-era influence networks. We investigated their social network metrics, degree distribution and connectivity. 
An influence pattern throughout all eras was that the internal impact of any era was higher than its external impact. The vast majority of scholars influenced scholars of their own period (= within-era influence) with an relatively stable average out-degree. There were only few instances of reciprocity.
When accumulating eras, the max. degrees drastically increased. However, over all eras, the maximum out-degree stayed greater than the maximum in-degree. 
In inter-era influence networks, each era influenced most its consecutive one, and additionally the Contemporary period. Exception to this rule was a spike in the absolute links of antique influences on the Early Modern Period, representing the increased reception of antique scholars during the Renaissance. However, proportionally Antiquity's influence on Early Modernity was as high as on the Middle Ages, which reasserts the shift in historical research that the Renaissance thinkers did not ``re-discover" Antiquity, but that medieval scholars also received it \cite[p. 3-4]{fejfer:2003}.


With a longitudinal perspective, we can add a more pronounced view on who the most influential intellectuals were. The scholar with the highest out-degree over all periods on contemporaries (= within-era) was Modern age scholar Friedrich Nietzsche. Plato in Antiquity, Avicenna in the Middle Ages, John Locke in the Early Modern period, Johann Goethe in the Transition period and Vladimir Nabokov in Contemporary were the most influential on the contemporaries of their respective periods. 

When accumulating eras, the most influential intellectuals of an era change: there, Plato was the most influential for Antiquity and the Middle Ages, Aristotle for the Early Modern and Transitioning period, Immanuel Kant for Modern Age. In the Contemporary period, and therefore for the complete network of intellectuals, it was Karl Marx. 

In the inter-era network analysis, Transitioning period scholar Karl Marx had the highest out-degree over all times to the Contemporary age. Second places over all time took Modern intellectuals Friedrich Nietzsche and Martin Heidegger on the Contemporary period.

We constructed the longitudinal influence power of intellectuals based on the average of their influences on eras, which favours consistency of influence.  Here, again Aristotle, Thomas Aquinas, William Shakespeare, Karl Marx, Friedrich Nietzsche and Vladimir Nabokov were the most consistently influential intellectuals of their respective periods. The highest influences had Nietzsche, Nabokov, and Marx.

In terms of knowledge brokering, we could identify Coordinator, Gatekeeper, Representative and Liasion knowledge brokers, whom we interpreted as passing influence between and within eras. We found scholars with all four different brokerage roles were medieval scholar Thomas Aquinas, Early Modern polygraph Gottfried Leibniz, Georg Hegel of the Transitioning Period, and the Modern philosopher Martin Heidegger. 

This study of the longitudinal patterns of influence such is suited to further the insights on the interconnections of influence of thinkers, and the dynamics of eras alike.


Therefore we plan to study the evolution of communities in these accumulated networks in future work.
In addition, we like to compare this YAGO network of intellectual influence with a more detailed network of scholars based on the main books on intellectual history, in order to establish their differences and insights on this field.

\bibliographystyle{plain}
\bibliography{main}

\end{document}